\documentclass[conference]{IEEEtran}
\IEEEoverridecommandlockouts
\usepackage{cite}
\usepackage{amsmath,amssymb,amsthm,bm,mathrsfs,amsfonts,dsfont}
\DeclareMathOperator*{\argmax}{arg\,max} 
\usepackage{algorithmic}
\usepackage{graphicx}
\usepackage{textcomp}
\usepackage{xcolor}
\usepackage{multirow}
\usepackage{tabularx}
\usepackage{tcolorbox}
\usepackage{caption}
\usepackage{subcaption}
\usepackage[colorlinks,bookmarksopen,bookmarksnumbered,citecolor=blue,urlcolor=blue]{hyperref}
\usepackage[linesnumbered,ruled,vlined]{algorithm2e}
\SetKwInput{KwInput}{Input}
\SetKwInput{KwOutput}{Output}
\allowdisplaybreaks
\newtheorem{remark}{Remark}
\def\BibTeX{{\rm B\kern-.05em{\sc i\kern-.025em b}\kern-.08em
    T\kern-.1667em\lower.7ex\hbox{E}\kern-.125emX}}
\begin{document}

\title{Joint Resource and Admission Management for Slice-enabled Networks
}

\author{\IEEEauthorblockN{Sina Ebrahimi*, Abulfazl Zakeri*, Behzad Akbari*, Nader Mokari*}
\IEEEauthorblockA{\textit{*Faculty of Electrical and Computer Engineering (ECE)} \\
\textit{Tarbiat Modares University}\\
Tehran, Iran \\
\{sina.ebrahimi, abolfazl.zakeri, b.akbari, nader.mokari\}@modares.ac.ir
}}


\maketitle

\begin{abstract}
Network slicing is a crucial part of the 5G networks that communication service providers (CSPs) seek to deploy. By exploiting three main enabling technologies, namely, software-defined networking (SDN), network function virtualization (NFV), and network slicing, communication services can be served to the end-users in an efficient, scalable, and flexible manner. To adopt these technologies, what is highly important is how to allocate the resources and admit the customers of the CSPs based on the predefined criteria and available resources. In this regard, we propose a novel joint resource and admission management algorithm for slice-enabled networks. In the proposed algorithm, our target is to minimize the network cost of the CSP subject to the slice requests received from the tenants corresponding to the virtual machines and virtual links constraints. Our performance evaluation of the proposed method shows its efficiency in managing CSP's resources.

\end{abstract}


\section{Introduction}
With the rise of the fourth industrial revolution in recent years, telecommunication industry leaders e.g. international telecommunication union (ITU) among some tier 1 service providers have tried to open the room for vertical industries by innovations in defining new services through the fifth generation of mobile networks (5G) paradigm.
ITU has defined some {stringent} contradicting requirements on reliability, latency, and throughput for 5G \cite{series2015imt}. 

To satisfy the aforementioned diverse requirements of IMT-2020 \cite{series2015imt}, the next generation mobile networks alliance (NGMN) has come up with the network slicing concept in which multiple logical networks could be deployed on a single physical network \cite{alliance20155g,alliance2016description}. The main target of network slicing for a communication service providers (CSPs) is to prepare an intelligent and flexible network which can support three (or more) conflicting types of services such as enhanced mobile broadband (eMBB), massive machine-type communications (mMTC), and ultra-reliable low-latency communications (URLLC) \cite{series2015imt}. The main goal of this paper is allocating the resources of {each} CSP to the received slice requests of its tenants by leveraging {network function virtualization (NFV) and software-defined networking (SDN)}.

\subsection{Related Works}
Herein, we divide the related works into two major categories in terms of network architecture and resource allocation in network slicing.
\subsubsection{Network Architecture}
In \cite{rost2016mobile}, the authors introduce a network slicing architecture towards 5G. In \cite{nikaein2015network}, the network store framework is introduced, in which tenants can provide virtual network functions (VNFs) for their slice and the CSP can manage all slices using it. The authors in \cite{samdanis2016network}, discuss the evolution of the network sharing concept towards an architecture of network slicing. They propose a slice broker architecture to manage the CSPs slices which belong to different tenants. 
In \cite{richart2016resource,zhang2017network,afolabi2018network,ordonez2017network}, SDN, NFV, and orchestration are introduced as the most relevant technologies to elevate the architecture of network slicing.
In this paper, the high-level architecture is inspired by the infrastructure sharing {aspect} of network slicing with an eye on QoS of different tenants \cite{richart2016resource,ordonez2017network}. Moreover, the multi-tenancy framework in this paper is inspired by \cite{ordonez2017network}. 
\subsubsection{{Resource Allocation in Network Slicing}}
\cite{su2019resource} is a comprehensive survey about {resource allocation} in network slicing. In \cite{sciancalepore2017slice}, the problem of network slice broker is formulated in order to provide a joint internet of things (IoT)/slice broker orchestration scheme. This approach, known as "Slice as a Service" (SlaaS) \cite{sciancalepore2017slice}, inherently improves the resource utilization rate. An admission and allocation problem of slice requests in order to maximize the revenue of CSP and satisfy the tenants' service requirements is discussed in \cite{bega2017optimising}. In \cite{han2018slice}, the authors take the SlaaS approach and {develop} an online genetic-based optimizer that approaches toward the ideal slicing strategy with {the} maximized long-term network utility. 

In \cite{leanh2016resource}, the authors introduce a {resource allocation} problem by adapting a matching game for a mobile virtual network operator, who buys physical resources from the CSP and bundle them into virtual resources called slices. \cite{sattar2019optimal} formulates a {mixed-integer linear programming (MILP)} for optimal allocation of a slice in 5G core networks. An integer linear program for offline mobile network slice embedding, focusing on {resource allocation} and virtual node and link mapping is proposed in \cite{fendt2018network}, to maximize the weighted sum of all embedded slices in the physical network. In \cite{wang2017resource}, the authors study the network slice dimensioning problem with resource pricing policy by exploring the relationship of the CSP (which is the slice provider) and the tenants (which are slice customers). A dynamic resource adjustment algorithm based on a reinforcement learning approach from each tenant's perspective is discussed in \cite{kim2019reinforcement} aiming to maximize the profit of CSP. Moreover, a network slicing framework, including admission control, resource allocation, and user dropping is presented in \cite{caballero2018network}, which use a game-theoretic approach to solve the problem.

Although there are some works in the literature about resource allocation in network slicing, there are still many challenges in this area that are not completely solved. For example, Resource pricing, working on end-to-end latency, examining the roles of the new players, and providing admission control mechanisms are some important challenges about resource allocation for network slicing. Resource allocation frameworks for network slicing have mostly failed to consider the relationship of the CSP with its tenants, e.g., third-parties, and vertical industries thoroughly. Moreover, despite envisioning slice request-based architectures in \cite{han2018slice} and \cite{wang2017resource}, they do not propose any admission control mechanisms to ensure the feasibility of their problems.

\subsection{Contributions}
In this paper, we propose a new SlaaS framework for a CSP which intends to allocate its own resources to the slice requests received from its tenants with formulating a novel MILP problem. The main contributions of this paper are presented as follows:
\\$\bullet~$Our proposed {joint resource allocation (JRA)} method minimizes the power consumption cost of the cloud nodes alongside the bandwidth consumption cost of links. This way, our proposed method prefers less turned-on cloud nodes. 
\\$\bullet~$
Our system model is more realistic than the related works, because tenants' slice requests affect the CSP's network and normally this impact is neglected {in the literature}. This framework allows the tenants to orchestrate their own slice and serve their own customers\footnote{Most tenants such as MVNOs or vertical industries (e.g., automotive and device manufacturing industry companies) know the requirements of their requested network and instantiate their own VNFs on their slice.}.  
\\$\bullet~$ To solve the proposed optimization problem, we devise two different methods and analyze them from different aspects. Moreover, to overcome the infeasibility of {these methods,} which is because of the limited amount of physical resources, we propose a novel admission control mechanism. Our proposed joint {admission control} mechanism (AC-JRA) gains about {$46\%$} on average in comparison to the disjoint admission control mechanism (AC-DRA).


\subsection{Organization}The rest of this paper is organized as follows. Section \ref{system-model} introduces the system model and categorizes the constraints of the problem. {In Section \ref{problem-formulation}, we first describe the formulation of the proposed joint optimization problem, and then linearize one of its constraints. Later on Section \ref{problem-formulation}, we propose an {admission control} mechanism. Moreover, a disjoint formulation of the problem is presented at the end of Section \ref{problem-formulation}}. The simulation results are presented in Section \ref{experimental-analysis}.
 Finally, we conclude the paper in Section \ref{conclusion}.



\section{System Model}\label{system-model}
{In this paper, we suppose a single CSP and multiple tenants, where the CSP serves the slice requests of these tenants.}
A typical illustration of the considered system model is depicted in Fig. \ref{architecture}. The details of our system model are as follows.

\subsection{CSP's Network}
\begin{figure}[!h]
	\centering
	\includegraphics[width=\columnwidth]{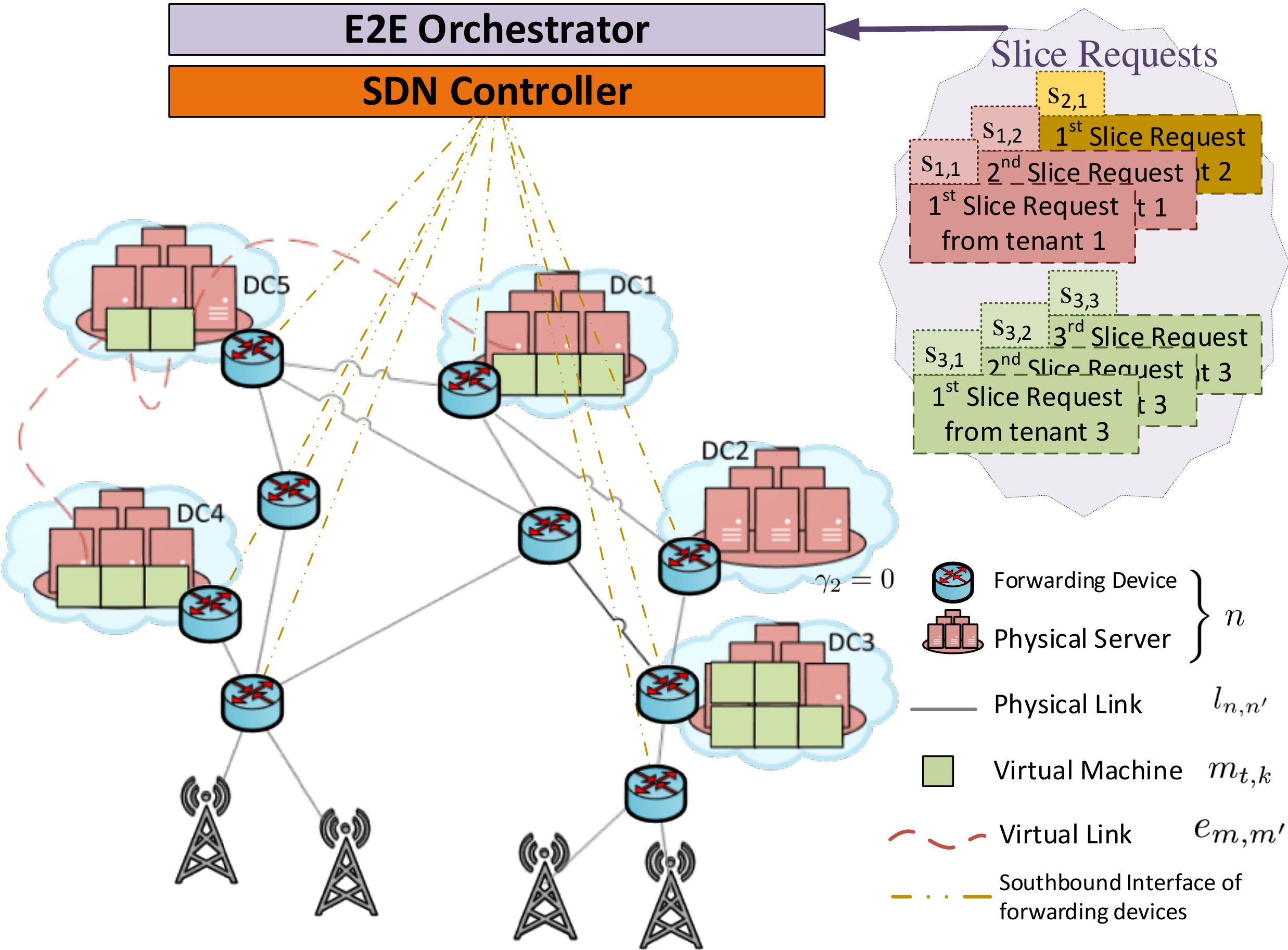}
	\caption{Network architecture of the CSP in our system model.}\label{architecture}
\end{figure}

We consider {the} CSP's network as a graph G($\mathcal{N},\mathcal{L}$), where $\mathcal{N}$ is the set of cloud nodes (physical servers/forwarding devices)\footnote{We use the terms node, physical node, data center (DC) 
and cloud node interchangeably.} and $\mathcal{L}=[l_{u,u'}]$ is the set of physical links, in which $l_{u,u'}=1$, if cloud nodes $u$ and $u'$ are connected, otherwise is $0$. We also assume that {the} CSP's cloud nodes are all NFV-enabled and the connectivity of all these {cloud nodes} is controlled by a logically-centralized SDN controller (Fig. \ref{architecture}). 

We indicate the total resource capacity of cloud node $n\in\mathcal{N}$ as $\textbf{r}_{n} = [r_{n}^{\text{Com}}\,\, r_{n}^{\text{Mem}}~\, r_{n}^{\text{Sto}}]$, where $r_{n}^{\text{Com}}$, $r_{n}^{\text{Mem}}$, and $r_{n}^{\text{Sto}}$ indicate the computational, memory, and storage capacities of $n$, respectively.

Each physical link $l_{n,n'}\in \mathcal{L}$ has the limited bandwidth of $BW^{l_{n,n'}}$. Moreover, since {the} cloud nodes are distributed geographically, there is a considerable propagation delay between them denoted by $\tau^{l_{n,n'}}$.

\subsection{Slice Requests}
We consider $\mathcal{T}=\{1,\dots,T\}$ as the set of all tenants and $\mathcal{K_\text{t}}=\{1,\dots,K_t\}$ as the set of tenant $t$'s slices in CSP's network. $K_t$ is the number of slice requests of tenant $t$. Each tenant $t$ requests its $k^\text{th}$ specific slice denoted by $s_{t,k}\in\mathcal{S}$ ($\mathcal{S}$ is the set all slice requests) with a specified set of virtual machines (VMs) and virtual links (VLs) indicated by $s_{t,k}=(\mathcal{{M}}_{t,k}, \mathcal{{E}}_{t,k})$ where $\mathcal{{M}}_{t,k}$ and $\mathcal{{E}}_{t,k}$ are the set of $s_{t,k}$'s requested VMs and VLs, respectively.  
We denote $\boldsymbol{\phi}_{m_{t,k}}=[\phi_{m_{t,k}}^{\text{Com}}\,\, \phi_{m_{t,k}}^{\text{Mem}}~\, \phi_{m_{t,k}}^{\text{Sto}}],\,m_{t,k}\in\mathcal{{M}}_{t,k}$ as the vector of the requested computing, memory and storage resources for the $m^\text{th}$ VM in $s_{t,k}$. 

Hereafter, we drop subscript $t,k$ from $m$ and $m'$ for the sake of readability and convenience in the rest of this paper. We denote $\boldsymbol{\chi^{e_{m,m'}}}=[\varpi^{e_{m,m'}}, \tau^{e_{m,m'}}_{\max}] ,\,\forall m,m'\in\mathcal{{M}}_{t,k}$ as the vector of requested VLs between VM $m$ and $m'$, where $\varpi^{e_{m,m'}}$ is the requested data rate between $m$ and $m'$ in $s_{t,k}$ and $\tau^{e_{m,m'}}_{\max}$ is the maximum tolerable delay between its two connected VMs $m$ and $m'$.

After $s_{t,k}$ arrives at the CSP's E2E orchestrator, {if there were enough resources left in the network}, it creates the slice templates and then calculates the cost based on the required resources. Then, the CSP creates $s_{t,k}$ by utilizing {the} specific cloud nodes and physical links that establish connections between those nodes\footnote{SDN controller must establish the physical connection for the requested VLs of each slice, which their paths are determined by the considered E2E orchestrator after solving the resource allocation problem.}. This relationship between the CSP and its tenants is similar to the business model introduced by 5GPPP \cite{5g2017view}.

\subsection{{Allocating Resources of the Cloud Nodes to Virtual Machines}}
The CSP creates VM $m$ by utilizing existing cloud node resources. We define the decision variable $\xi_{m}^n$, which is set to $1$ if VM $m$ is created on $n$ by hypervisor for slice $s_{t,k}$, otherwise is $0$.
To ensure that the sum of allocated VM resources in a cloud node do not exceed its resources, we have the following constraint:
\begin{align}
\label{vmconstraint}
\sum_{t\in\mathcal{{T}}} \sum_{k\in\mathcal{K_\text{t}}} \sum_{m \in \mathcal{M}_{t,k}} \xi_{m}^n .\phi_{m} \preceq \mathbf{r}_{n} \qquad \forall n \in \mathcal{N}.
\end{align}

\subsection{Power Usage of the Cloud Nodes}
We define the variable $\gamma_{n}$ which is set to $1$ if node $n$ is on, otherwise is $0$.
To verify which nodes are turned on, we have the following constraint as \cite{zakeri2019energy}: 
\begin{align}
\label{onoffconstraint}\xi_{m}^n\le \gamma_{n},~~\forall n\in\mathcal{N},~\forall m \in \mathcal{M}_{t,k}, \forall{t}\in\mathcal{T},\,\forall{k}\in\mathcal{K_\text{t}}.
\end{align}
We define the power consumed by {node} $n$ inspired by \cite{gao2013quality,beloglazov2012energy,dayarathna2016data}:
\begin{equation}\label{dc-power}
P_{n}= (P^{\text{max}}_n - P^{\text{idle}}_n) U^{\text{Com}}_n + \gamma_{n} P^{\text{idle}}_n ,
\end{equation}
where $P^{\text{idle}}_n$ and $P^{\text{max}}_n$ are the average power values when node $n$ is idle and is fully utilized, respectively. Moreover, we define the CPU utilization of {node} $n$ as follows \cite{gao2013quality,beloglazov2012energy}:
\begin{equation}\label{cpu-util}
U^{\text{Com}}_n = \frac{\sum_{t\in\mathcal{{T}}}\sum_{k\in\mathcal{K_\text{t}}}\sum_{m \in \mathcal{M}_{t,k}} \xi_{m}^n .\phi^\text{Com}_m}{r^\text{Com}_n}
\end{equation}

\subsection{Determining the Physical Path for Virtual Links}
There may be multiple physical paths between two cloud nodes. 
Therefore, we should choose the best physical path for each virtual link. 
 We denote  $\mathcal{B}_{n,n'}=\{1, \cdots, b, \cdots, B_{n,n'}\}$ as the set of total possible physical paths, where $B_{n,n'}$ is the total number of physical paths between cloud nodes $n$ and $n'$.  We also denote the $b^{\rm th}$ physical path between cloud nodes $n$ and $n'$ as $p_{n,n'}^b$. We define the decision variable $\pi_{p_{n,n'}^b}^{e_{m,m'}}, m \neq m'$ where it is set to $1$ if VL $e_{m,m'}$ is sent over path $p_{n,n'}^b$, otherwise is set to $0$.
To determine whether physical link $l_{u,u'}$ contributes in physical path $p_{n,n'}^b$, the {binary} indicator $I^{l_{u,u'}}_{p_{n,n'}^b}$ is defined.

Moreover, the set of all physical links that contribute in a path $p_{n,n'}^b$  is defined as $\mathcal{L}_{p_{n,n'}^b}=\left\{l_{u,u'}\in \mathcal{L}~\Big|~I^{l_{u,u'}}_{p_{n,n'}^b}=1\right\}$.
To ensure that the path between nodes $n$ and $n'$ is established for the connectivity of VMs $m$ and $m'$, we introduce the following constraints:
\begin{align}
\label{vlpathconstraint}
\sum_{n\in\mathcal{{N}}} \sum_{n'\in\mathcal{{N}}} \sum_{b\in\mathcal{{B}}_{nn'}} \pi^{e_{m,m'}}_{p^b_{n,n'}} = {1}  ,\,
\end{align}

\begin{align}
&\pi^{e_{m,m'}}_{p^b_{n,n'}} = \xi_{m}^n \xi_{m'}^{n'} ,\,\label{vlpathconstraint2} \forall e_{m,m'} \in \mathcal{E}_{t,k},\forall{t}\in\mathcal{T},\,\forall{k}\in \mathcal{K_\text{t}}  , n \neq n'. 
\end{align}
Constraint \eqref{vlpathconstraint2} ensures that the right physical path is chosen between cloud nodes $n$ and $n'$, which are the hosts of VMs $m$ and $m'$.

\subsection{Bandwidth {Limitations}}
The aggregated rates of all VLs belonging to all of the slices that pass a physical link should not exceed its bandwidth. This can be ensured if
\begin{align}
\nonumber
&\sum_{t\in\mathcal{{T}}}\sum_{k\in\mathcal{K_\text{t}}}\sum_{e_{m,m'} \in \mathcal{E}_{t,k}} \sum_{n\in\mathcal{{N}}} \sum_{n'\in\mathcal{{N}}} \sum_{b\in\mathcal{{B}}_{nn'}} I^{l_{u,u'}}_{p_{n,n'}^b} \pi^{e_{m,m'}}_{p^b_{n,n'}} \varpi^{e_{m,m'}} \le\\&\qquad\qquad\qquad\qquad\label{vl_bw_constraint}BW^{l_{u,u'}} ,\, \forall l_{u,u'}\in\mathcal{L}_{p_{n,n'}^b}, n \neq n' 
\end{align}
holds. We 
calculate the overall traffic of all VLs belonging to all the slices that pass through these physical links and determine the cost of passing this traffic. The overall cost of bandwidth consumption is calculated as
\begin{align}
\small
\label{bw_consumption}
\beta= \sum_{t\in\mathcal{{T}}}\sum_{k\in\mathcal{K_\text{t}}}\sum_{e_{m,m'} \in \mathcal{E}_{t,k}} \sum_{b\in\mathcal{{B}}_{nn'}} \sum_{l\in\mathcal{L}_{p_{n,n'}^b}} I^{l_{u,u'}}_{p_{n,n'}^b} \pi^{e_{m,m'}}_{p^b_{n,n'}} \psi^{l_{u,u'}} \varpi^{e_{m,m'}},
\end{align}
where $\psi^{l_{u,u'}}$ is the cost of transmitting 1bps traffic on $l_{u,u'}$.

\begin{remark}\label{remark_samenode}
	For the scenario in which $m$ and $m'$ are mapped to the same cloud node ($n=n'$), we assume that we have the physical connection between these VMs with much higher data rate and negligible delay. 
\end{remark}

\subsection{{Maximum Tolerable Delay of Virtual Links}}
To ensure that we can guarantee tenant's maximum tolerable propagation delay ($\tau^{e_{m,m'}}_{\max}$) between its VLs, we have the following constraint:
\begin{align}\nonumber
\sum_{l_{u,u'} \in \mathcal{L}} I^{l_{u,u'}}_{p_{n,n'}^b} \pi^{e_{m,m'}}_{p^b_{n,n'}} \tau^{l_{u,u'}} &\le \tau^{e_{m,m'}}_{\max} ,\,\forall e_{m,m'}\in\mathcal{E}_{t,k} ,\,\\& \qquad \qquad \forall{t}\in\mathcal{T},\,\forall{k}\in\mathcal{K_\text{t}}.
\label{vldelayconstraint}
\end{align}
It is noteworthy that because the CSP does not have information on the applications running on top of tenants' VMs and how much data will arrive at each VM, we cannot calculate the execution delay in this case. Also, since the amount of data passing the physical links is unknown and the CSP only guarantees the requested data rates of the tenants, the transmission delay (neither on the radio nor the network side) cannot be computed either.

{To make the paper more readable, Table \ref{table-notation} provides the main notation used in this paper.}
\begin{table}[h]
	\renewcommand{\arraystretch}{1.05}
	\centering
	\scriptsize
	\caption{{Main Notation}}
	\label{table-notation}
	\begin{tabular}{>{\color{black}}c |>{\color{black}}c}

		\textbf{Notation}& \textbf{Definition}\\\hline
		$\mathcal{T}$&Set of tenants\\ \hline
		$\mathcal{K_\text{t}}$ &Set of tenant $t$'s slices\\ \hline
		$s_{t,k}$ &The $k^\text{th}$ slice of tenant $t$\\ \hline
		$\mathcal{{M}}_{t,k}$ &Set of $s_{t,k}$'s requested VMs\\ \hline
		$\mathcal{{E}}_{t,k}$ &Set of $s_{t,k}$'s requested VLs\\ \hline
		$\boldsymbol{\phi}_{m_{t,k}}$ &The vector of requested virtual resource capacities for $m_{t,k}$\\ \hline
		$\textbf{r}_{n}$ &The vector of physical resource capacities of cloud node $n$\\ \hline
		$\boldsymbol{\chi^{e_{m,m'}}}$&The vector of requested VLs between $m_{t,k}$ and $m'_{t,k}$\\\hline
		$\varpi^{e_{m,m'}}$ &Tenant $t$'s requested data rate between $m_{t,k}$ and $m'_{t,k}$\\ \hline
		$\tau^{e_{m,m'}}_{\max}$ &\begin{tabular}{c}Tenant $t$'s requested maximum tolerable propagation\\delay between $m_{t,k}$ and $m'_{t,k}$\end{tabular}\\ \hline
		$\mathcal{N}$ &Set of CSP's cloud nodes/physical servers\\\hline
		$\mathcal{L}$ &Set of CSP's physical links\\\hline
		$l_{n,n'}$ &Physical link between nodes $n$ and $n'$\\\hline
		$BW^{l_{n,n'}}$ &The bandwidth of the link $l_{n,n'}$\\\hline
		$\psi^{l_{u,u'}}$ &The cost of transmitting 1 bps on the link $l_{u,u'}$\\\hline
		$I^{l_{u,u'}}_{p_{n,n'}^b}$&\begin{tabular}{c}Indicator determining that physical link $l_{u,u'}$\\contributes in the $b^{\text{th}}$ path between $n$ and $n'$\end{tabular}\\\hline				
		$\gamma_{n}$&Decision variable is set to 1 if cloud node $n$ is switched on\\\hline
		$\xi_{m}^n$&\begin{tabular}{c}Decision variable is set to 1 if VM $m_{t,k}$\\is embedded on cloud node $n$\end{tabular}\\\hline
		$\pi_{p_{n,n'}^b}^{e_{m,m'}}$&\begin{tabular}{c}Decision variable is set to 1 if VL $e_{m,m'}$ is\\ mapped on the $b^{\text{th}}$ path between $n$ and $n'$\end{tabular}\\\hline
		$\Upsilon, \zeta$ & \begin{tabular}{c}Cost weight functions for balancing the power usage cost of\\all cloud nodes and the bandwidth consumption of VLs\end{tabular}\\\hline
		$C_{\text{Total}}$ &Total cost of SP's network\\
	\end{tabular}
\end{table}

\section{The Proposed Resource Allocation and Admission Control of Slice Requests} \label{problem-formulation}
In this section, we first formulate the joint resource allocation problem (JRA) and then linearize one of its constraints. Then, in order to make sure the aforementioned problem would not be infeasible, we propose an admission control mechanism for the received slice requests which is solved before the proposed JRA. Moreover, we introduce disjoint VM allocation (DMA) and disjoint VL allocation (DLA) problems, i.e., DRA with their corresponding admission control mechanisms to evaluate our JRA method at the end of this section. 

\subsection{The Proposed Joint Resource Allocation Problem}
In this subsection, we formulate the proposed joint VM and VL placement. 
Aiming to minimize the total network cost for the CSP, we define the total cost function based on the minimization of active cloud nodes and reducing the cost of bandwidth consumption. The overall cost of resource allocation consists of two components: 1) bandwidth consumption cost, which relates to the power consumption of forwarding devices as well as the specific operational expenditure (OPEX) on diverse physical links, 2) the cost of consumed power of all turned-on cloud nodes. Therefore, the overall cost function can be stated as

\begin{align}\label{objective_function}  C_{\text{Total}}(\boldsymbol{\pi},\boldsymbol{\gamma},\boldsymbol{\xi})=\zeta \beta +\Upsilon \sum_{n\in\mathcal{N}} P_{n} ,\, 
\end{align} where $\boldsymbol{\pi},\boldsymbol{\gamma},\boldsymbol{\xi}$ are the sets of all $\pi^{e_{m,m'}}_{p^b_{n,n'}}$, $\gamma_{n}$, $\xi_{m}^n$, respectively. Moreover, $\zeta$ and $\Upsilon$ are scaling factors for translating bandwidth and power consumption into cost. Hence, the joint VM and VL placement optimization problem for the management of the requested slices by the CSP can be written as

\begin{equation}\label{joint_optimization} 
\begin{array}{ll}
\mathop{\min}\limits_{{\bm \pi},{\bm \gamma},{\bm \xi}}& C_{\text{Total}}(\boldsymbol{\pi},\boldsymbol{\gamma},\boldsymbol{\xi})\\
\text{subject to}&  \text{C1:}\quad \eqref{vmconstraint}\\
&\text{C2:}\quad \sum_{n \in \mathcal{N}} \xi_{m}^n = 1,\quad \forall t,k,m\\
&\text{C3:}\quad \eqref{onoffconstraint}\\
&\text{C4:}\quad \eqref{vlpathconstraint}\\
&\text{C5:}\quad \eqref{vlpathconstraint2}\\
&\text{C6:}\quad \eqref{vl_bw_constraint}\\
&\text{C7:}\quad \eqref{vldelayconstraint}\\
&\text{C8:}\quad \gamma_n \in \{0,1\}, \quad \forall n\\
&\text{C9:}\quad \xi_m^n \in \{0,1\}, \quad \forall m,n\\
&\text{C10:}\quad \pi^{e_{m,m'}}_{p^b_{n,n'}} \in \{0,1\},\quad \forall e,b,n.\\
\end{array}
\end{equation}

Constraint C1 makes sure that all VMs are hosted by their corresponding cloud nodes without violation in computational, memory, and storage capacities of the cloud nodes, respectively. Constraint C2 guarantees that each VM will be placed only in one cloud node. Constraint C3 ensures that we host VMs on turned-on cloud nodes. C4 and C5 make sure that each VL is only traversing one physical path and its path is between the cloud nodes hosting its corresponding VMs. Moreover, constraints C6 and C7 are related to bandwidth and delay limitations, respectively. Finally, constraints C8-C10 assure that decision variables of the problem are binary. 

\subsection{Solution to the JRA problem}
The introduced joint problem \eqref{joint_optimization} is an integer non-linear programming problem due to the constraint C5. Since there are effective integer linear programming (ILP) solvers i.e., MOSEK, we should linearize C5 first. Influenced by \cite{addad2018towards}, we replace the $\xi_{m}^n \xi_{m'}^{n'}$ with an auxiliary variable $\theta_{m,m'}^{n,n'}$. Furthermore, to ensure that the equality $\theta_{m,m'}^{n,n'}=\xi_{m}^n \xi_{m'}^{n'}$ holds, we replace C5 in problem \eqref{joint_optimization} with inequalities below:

\begin{equation}\label{c5_linear} 
\small
\begin{array}{ll}
&\text{C5-a:}\quad \pi^{e_{m,m'}}_{p^b_{n,n'}} = \theta_{m,m'}^{n,n'} ,\\
&\text{C5-b:}\quad \theta_{m,m'}^{n,n'} \leq \xi_{m}^n +1-\xi_{m'}^{n'},\\
&\text{C5-c:}\quad \xi_{m}^n \leq \theta_{m,m'}^{n,n'} +1-\xi_{m'}^{n'},\\
&\text{C5-d:}\quad \theta_{m,m'}^{n,n'} \leq \xi_{m'}^{n'}, \qquad
\forall e_{m,m'} \in \mathcal{E}_{t,k},  \\
&\text{}\qquad\qquad \forall m,m' \in \mathcal{M}_{t,k}, \forall n,n' \in \mathcal{N},{t}\in\mathcal{T},\,{k}\in \mathcal{K_\text{t}}  , n \neq n'.\\
\end{array}
\end{equation}

\noindent{By applying these constraints to \eqref{joint_optimization} instead of C5, we make sure that both $\xi_{m}^n$ and $\xi_{m'}^{n'}$ are not zero when $\pi^{e_{m,m'}}_{p^b_{n,n'}}$ is set to $1$.}


\subsection{Admission Control Mechanism for {the Proposed Joint Resource Allocation}}
In this paper, we apply a novel admission control mechanism in order to determine which slice requests are making the joint cloud nodes and links infeasible. 
Since resources in the cloud nodes and the links between them are limited (constraints \eqref{vmconstraint}, \eqref{vl_bw_constraint}, \eqref{vldelayconstraint}), if the virtual resources demanded by the slice requests are more than the whole network's capacity, the problem would be infeasible and no slice request can be served by the network. Whenever the admission control problem is solved and the sum of all elastic variables (defined in Table \ref{table-elastic-variables}) of the problem is more than zero, we figure out that the original problem (i.e., \eqref{joint_optimization}) would be infeasible. {In each round of our heuristic method (i.e., Alg. \ref{algorithm_AC}), if the sum of elastic variables (i.e., the objective function of \eqref{admission_control}) is not zero, we will find out which slice request is demanding more resources and reject it. We repeat this until the problem \eqref{joint_optimization} becomes feasible and then we solve it using ILP solvers. }

\begin{table}[h]
	\renewcommand{\arraystretch}{1.05}
	\centering
	\caption{Definition of the elastic variables in \eqref{admission_control}}
	\label{table-elastic-variables}
	\begin{tabular}{| c| l| l| }	
		\hline
		\textbf{Variable(s)}& \textbf{Defined to elasticize}& \textbf{Const.}\\\hline
		${\bm \sigma^{VM}}$& \begin{tabular}[c]{@{}l@{}}Computing, Memory, and Storage capacity \\of the cloud nodes\end{tabular}&\eqref{vmconstraint}\\ \hline
		$\sigma^{BW}$&Bandwidth of the physical links&\eqref{vl_bw_constraint}\\ \hline
		$\sigma^{\tau}$&Maximum tolerable latency of VLs&\eqref{vldelayconstraint}\\ \hline	
	\end{tabular}
\end{table}

In order to find the slice requests which cause infeasibility, we use the elasticization approach in which, the corresponding constraints are elasticized by defining some elastic variables that extend the bounds on constraints \cite{Chinneck2008feasibility,tajallifar2019qosaware}. 
The problem of AC-JRA can be written as
\begin{equation}\label{admission_control} 
\begin{array}{ll}
&\mathop{\min}\limits_{\bm \sigma} ({\bm \sigma^{VM}}+\sigma^{BW}+\sigma^{\tau})\\
& \text{subject to}\\& \text{C1-a:}\, \sum_{t}\sum_{k}\sum_{m} \xi_{m}^n .\boldsymbol{\phi}_{m}\preceq \mathbf{r}_{n}+ {\bm \sigma^{VM}_n},\\
&\text{C6-a:}\quad \sum_{t}\sum_{k}\sum_{e_{m,m'}} \sum_{n} \sum_{n'} \sum_{b} I^{l_{u,u'}}_{p_{n,n'}^b} \pi^{e_{m,m'}}_{p^b_{n,n'}} \varpi^{e_{m,m'}} \\&\qquad\qquad\qquad\qquad\qquad\qquad\qquad\quad\le BW^{l_{u,u'}} + \sigma^{BW}_{l_{u,u'}}, \\
&\text{C7-a:}\quad \sum_{l_{u,u'}} I^{l_{u,u'}}_{p_{n,n'}^b} \pi^{e_{m,m'}}_{p^b_{n,n'}} \tau^{l_{u,u'}} \le \tau^{e_{m,m'}}_{\max} + \sigma^{\tau}_{s_{t,k}}, \\
&\text{C2, C3, C4, C5-a, C5-b, C5-c, C5-d, C8, C9, C10.}
\end{array}
\end{equation}
In \eqref{admission_control}, the value of $\sigma^{VM}_{n}$ represents the difference of the sum of allocated computing, memory, and storage resources from a feasible fulfilled set of these resources in node $n$. The value of $\sigma^{BW}_{l_{u,u'}}$ represents the deviation of the sum of all requested data rates of VLs from the actual bandwidth of the physical link $l_{u,u'}$. Also, the value of $\sigma^{\tau}_{s_{t,k}}$ shows the difference of one (or more) VL(s) of the slice request $s_{t,k}$ from the actual propagation delay(s) of physical links.
Alg. \ref{algorithm_AC} provides broader information about our admission control mechanism.
\begin{algorithm}
	\small
	\DontPrintSemicolon
	\KwInput{$\mathcal{N}, \mathcal{L} , \boldsymbol{r} , \boldsymbol{BW} , \boldsymbol{\psi} , \boldsymbol{\tau} , \boldsymbol{I},$\\ \qquad \quad $\mathcal{S} , \mathcal{M} , \mathcal{E} , \boldsymbol{\phi} , \boldsymbol{\varpi} , \boldsymbol{\tau_{\max}}$}
	
	\Repeat{$\sum{({\bm \sigma^{VM}}+\sigma^{BW}+\sigma^{\tau})}=0$}
	{
		Solve \eqref{admission_control} according to $\mathcal{S}$
		
		\If{$\sum{\bm \sigma^{VM}} \ne 0$}
		{
			
			\uIf{$\sum_{n\in\mathcal{{N}}}{\sigma^{VM, Com}_n} \ne 0$}
			{
				\For{$t \leq T$ and $k \leq K_t$}
				{
					$SumUsedCom(s_{t,k})=\sum_{m \in \mathcal{M}_{t,k}} \phi_{m_{t,k}}^{\text{Com}}$
				}
				$s^{\star}_{t,k} = \argmax_{s_{t,k}} (SumUsedCom(s_{t,k}))$					
			}
			\uElseIf{$\sum_{n\in\mathcal{{N}}}{\sigma^{VM, Mem}_n} \ne 0$}
			{
				\For{$t \leq T$ and $k \leq K_t$}
				{
					$SumUsedMem(s_{t,k})=\sum_{m \in \mathcal{M}_{t,k}} \phi_{m_{t,k}}^{\text{Mem}}$
				}
				$s^{\star}_{t,k} = \argmax_{s_{t,k}} (SumUsedMem(s_{t,k}))$				
			}
			\uElseIf{$\sum_{n\in\mathcal{{N}}}{\sigma^{VM, Sto}_n} \ne 0$}
			{
				\For{$t \leq T$ and $k \leq K_t$}
				{
					$SumUsedSto(s_{t,k})=\sum_{m \in \mathcal{M}_{t,k}} \phi_{m_{t,k}}^{\text{Sto}}$
				}
				$s^{\star}_{t,k} = \argmax_{s_{t,k}} (SumUsedSto(s_{t,k}))$
			}		
			set all requested resources of $s^{\star}_{t,k}$ to zero and reject {this slice request}		
		}		
		\Else
		{
			\uIf{$\sum_{l_{u,u'} \in \mathcal{L}}{\sigma^{BW}_{l_{u,u'}}} \ne 0$}
			{
				\For{$t \leq T$ and $k \leq K_t$}
				{
					$SumUsedRate(s_{t,k})=\sum_{e_{m,m'} \in \mathcal{E}_{t,k}} \varpi^{e_{m,m'}}$
					
				}
				$s^{\star}_{t,k} = \argmax_{s_{t,k}} (SumUsedRate(s_{t,k}))$
			}
			\uElseIf{$\sum_{l_{u,u'} \in \mathcal{L}}{\sigma^{\tau}_{s_{t,k}}} \ne 0$}
			{				
				$s^{\star}_{t,k} = \argmax_{s_{t,k}} \sigma^{\tau}_{s_{t,k}}$
			}
			set all requested resources of $s^{\star}_{t,k}$ to zero and reject {this slice request}		
		}
		Update $\mathcal{S}$
	}
	\KwOutput{${\mathcal{S}}$}	
	\caption{Joint {admission control} (i.e., AC-JRA) of the received slice requests (Solving \eqref{admission_control})}
	\label{algorithm_AC}
\end{algorithm}

\subsection{Summary of the Proposed Joint Method}
\begin{figure}[!h]
	\centering
	\includegraphics[width=\columnwidth]{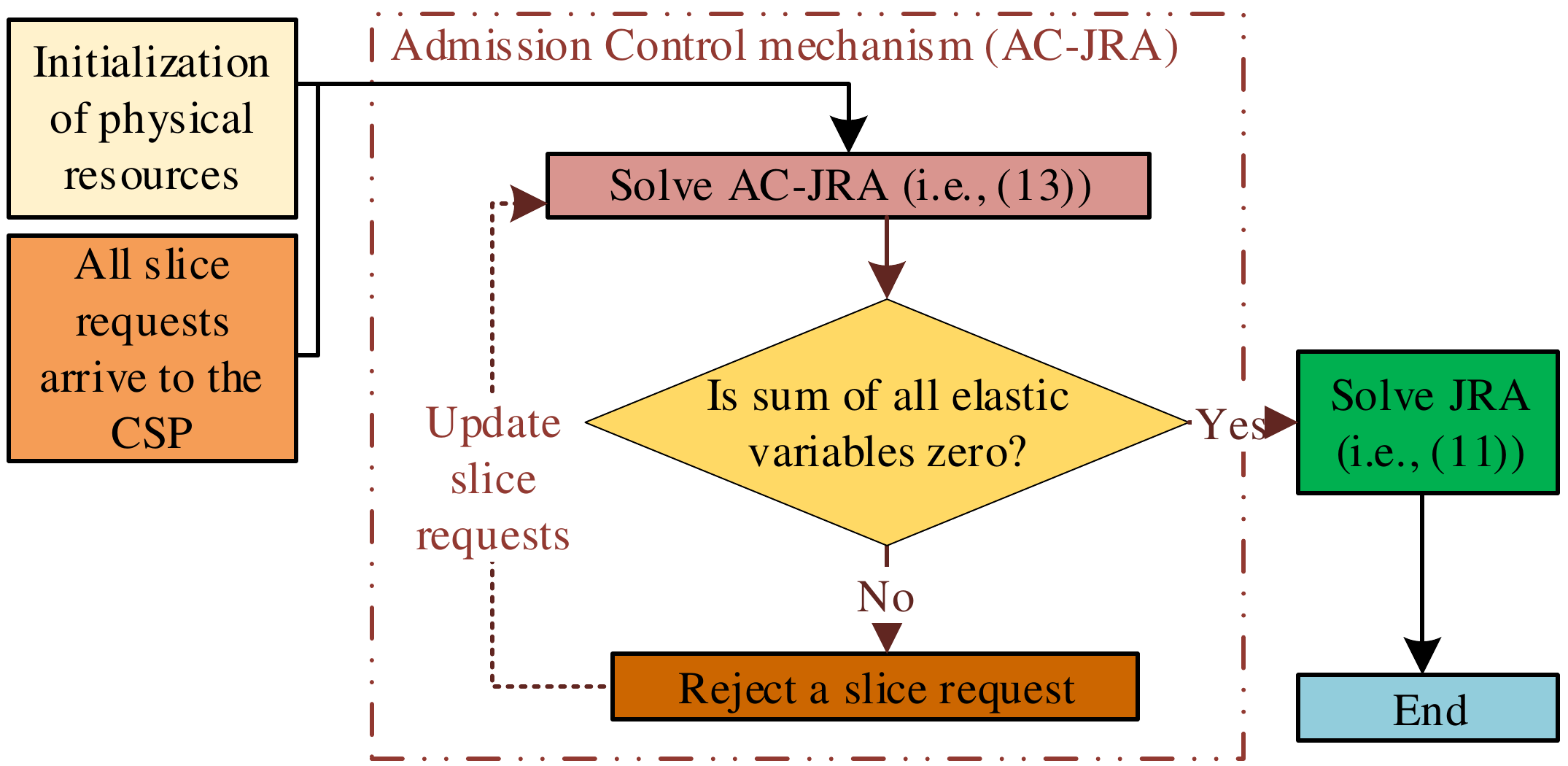}
	\caption{Simplified flowchart of {JRA}.}\label{flowchart}
\end{figure}

Herein, we summarize our proposed joint method which consists of {AC-JRA and JRA} for our network slicing framework. The proposed method can be illustrated in Fig. \ref{flowchart}. 
\subsection{{Disjoint Resource Allocation of Cloud Nodes and Links}}\label{disjoint-section}
\begin{figure}[!h]
	\centering
	\includegraphics[width=\columnwidth]{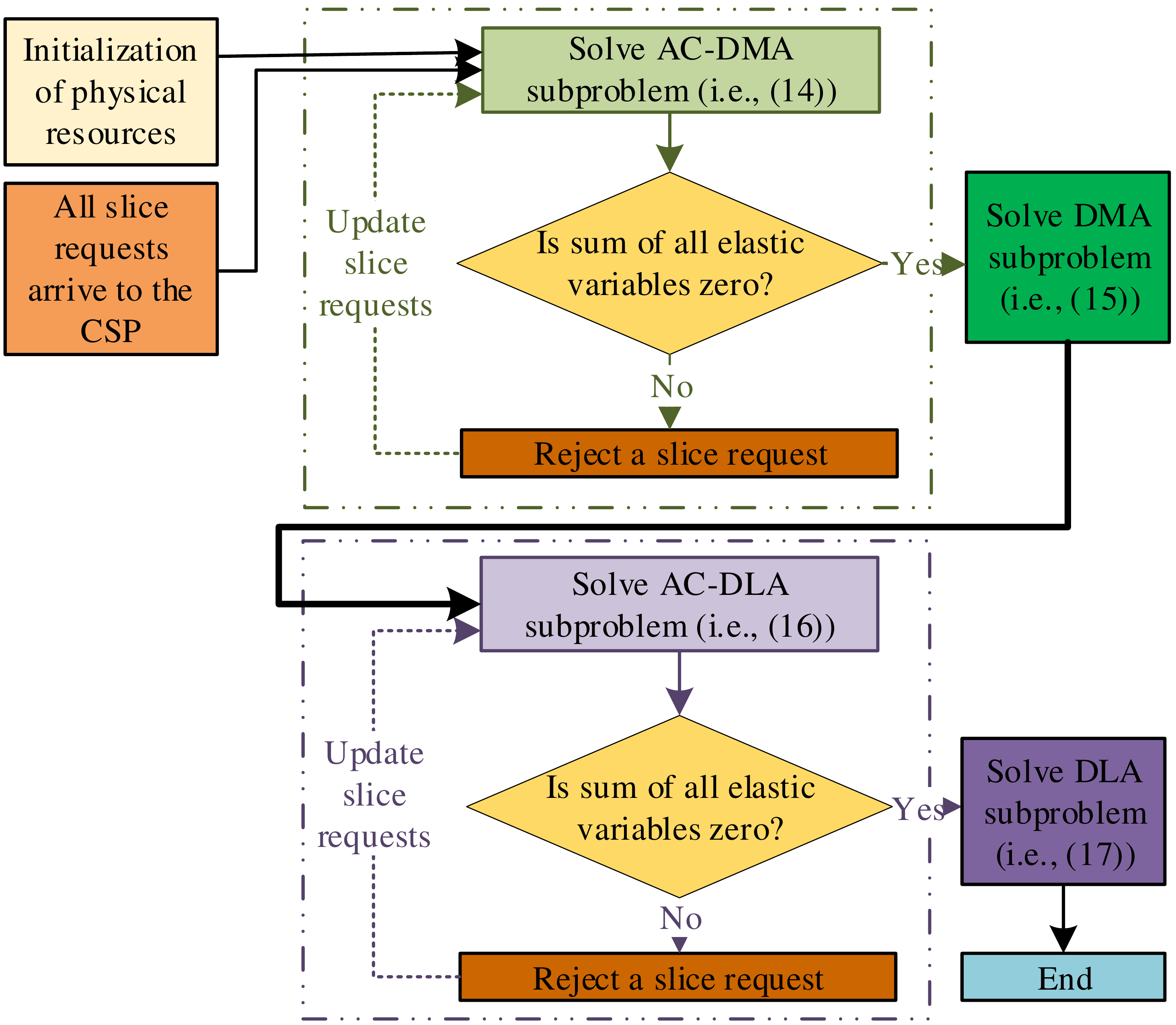}
	\caption{Simplified flowchart of DRA.}\label{disjoint-flowchart}
\end{figure}

In order to evaluate the proposed {JRA} problem (Fig. \ref{flowchart}), we advise a disjoint resource allocation (DRA) algorithm as a baseline of comparison. In this method, we divide {both problems of admission control and resource allocation} into nodes and links subproblems. The summary of the DRA method is shown in Fig. \ref{disjoint-flowchart}. {The proposed AC-DMA} can be written as
\begin{equation}\label{nodes_admission_control} 
\begin{array}{ll}
\mathop{\min}\limits_{\bm \sigma}& ({\bm \sigma^{VM}})\\
\text{subject to}&  
\text{C1-a, C2, C3, C8, C9.}\\
\end{array}
\end{equation}

\noindent Also, the {DMA subproblem} can be stated as
\begin{equation}\label{nodes_resource_allocation} 
\begin{array}{ll}
\mathop{\min}\limits_{{\bm \gamma},{\bm \xi}}& \Upsilon(\sum_{n\in\mathcal{N}} P_{n})\\
\text{subject to}&  
\text{C1, C2, C3, C8, C9.}\\
\end{array}
\end{equation}

\noindent After solving DMA subproblem, we solve the proposed AC-DLA knowing ${\bm \gamma}$ and ${\bm \xi}$ from \eqref{nodes_resource_allocation}:
\begin{equation}\label{links_admission_control} 
\begin{array}{ll}
\mathop{\min}\limits_{\bm \sigma}& (\sigma^{BW}+\sigma^{\tau})\\
\text{subject to}&  
\text{C4, C5, C6-a, C7-a, C10.}\\
\end{array}
\end{equation}
After the proposed AC-DLA, the last phase is to allocate link-related resources {(also known as DLA subproblem)} (e.g., $BW$ and $\tau$) which can be stated as:
\begin{equation}\label{links_resource_allocation} 
\begin{array}{ll}
\mathop{\min}\limits_{{\bm \pi}}& C_{\text{Total}}(\boldsymbol{\pi},\boldsymbol{\gamma},\boldsymbol{\xi})\\
\text{subject to}&  
\text{C4, C5, C6, C7, C10.}\\
\end{array}
\end{equation}
It is worth mentioning that {the} admission control mechanisms used in \eqref{nodes_admission_control} (i.e., AC-DMA) and \eqref{links_admission_control} (i.e., AC-DLA) are alike Alg. \ref{algorithm_AC}.

\section{Experimental Analysis} \label{experimental-analysis}
In this section, we evaluate the performance of the proposed joint (i.e., JRA) and disjoint (i.e., DMA and DLA) methods alongside their admission control mechanisms (i.e., AC-JRA vs. AC-DMA and AC-DLA). 
\subsection{Simulation Environment}
Herein, we explain the simulation environment used for the results in Section VI. Using a random topology with $4$ cloud nodes, we initialize the CSP's physical network. The values of the parameters related to the cloud nodes {(e.g., computational, memory, and storage resources)} are presented in the left part of Table \ref{table-node-parameters}. Moreover, the left-hand side of Table \ref{table-link-parameters} displays the values of the parameters related to the CSP's physical links {(e.g., propagation delay and bandwidth of links)}.
Moreover, the weights of node and link parts of the objective function \eqref{objective_function} are set to $\Upsilon=1$ and $\zeta=9\times10^{-5}$ experimentally {in order to balance the effects of these parts to each other}.
 
\begin{table}[]
	\centering
	\caption{{Cloud nodes simulation parameters}}
	\label{table-node-parameters}
	\begin{tabular}{|cc|cc|}
		\hline
		\multicolumn{1}{|c|}{\textbf{Resource Type}} & \textbf{Value} & \multicolumn{1}{c|}{\textbf{Requested Resource}} & \textbf{Value} \\ \hline
		$r^{Com}_n$ (MHz) & 7000 & $\phi_{m_{t,k}}^{\text{Com}}$ (MHz) & 1000 \\ \hline
		$r^{Mem}_n$ (GB) & 800 & $\phi_{m_{t,k}}^{\text{Mem}}$ (GB) & 64 \\ \hline
		$r^{Sto}_n$ (GB) & 2000 & $\phi_{m_{t,k}}^{\text{Sto}}$ (GB) & 120 \\ \hline
	\end{tabular}
\end{table}

For evaluating the effect of the number of slice requests on CSP's network, we increase the number of tenants from $1$ to {$16$}, keeping $k_t=1$ in all circumstances {(all tenants have only one slice request)}. We also keep the number of VMs in all slice requests the same ($|M|=3$). All VM requested resources of each slice request are shown in the right-hand side of Table \ref{table-node-parameters}. The values of the parameters related to the VL requests of each slice request is shown in the left part of Table \ref{table-link-parameters}. Note that the parameters that do not have the same $\max$ and $\min$ value presented in Table \ref{table-link-parameters} are set randomly between the aforementioned values.
It is worth mentioning that we use MOSEK solver in order to solve problems \eqref{joint_optimization}, \eqref{admission_control}, \eqref{nodes_admission_control}, \eqref{nodes_resource_allocation}, \eqref{links_admission_control}, and \eqref{links_resource_allocation}  \cite{mosek,cvx}. {Moreover, all simulation steps (including initialization, admission control mechanisms, and solving ILP problems with MOSEK toolbox) have been implemented in MATLAB software \cite{matlab} which is widely used to solve resource allocation problems \cite{sattar2019optimal,tajiki2019joint}.}

\begin{table}[]
	\centering
	\caption{CSP's network simulation parameters}
	\label{table-link-parameters}
	\begin{tabular}{|c|c|c|c|}
		\hline
		\textbf{CSP's Resource} & \textbf{In/Out} & \textbf{$\min$} & \textbf{$\max$} \\ \hline
		\multirow{2}{*}{\begin{tabular}[c]{@{}c@{}}$BW^{l_{n,n'}}$\\ (Kbps)\end{tabular}} & Intra & $10^7$ & $10^7$ \\ \cline{2-4} 
		& Inter & $9\times10^4$ & $1.9\times10^5$ \\ \hline
		\multirow{2}{*}{\begin{tabular}[c]{@{}c@{}}$\psi$\\ (\$)\end{tabular}} & Intra & $1$ & $1$ \\ \cline{2-4} 
		& Inter & $10^{-3} BW$ & $10^{-2} BW$ \\ \hline
		\multirow{2}{*}{\begin{tabular}[c]{@{}c@{}}$\tau^{l_{n,n'}}$\\ (ms)\end{tabular}} & Intra & $0$ & $0$ \\ \cline{2-4} 
		& Inter & $10^{-1}$ & $4$ \\ \hline
		\textbf{Requested Resource} & \multicolumn{3}{l|}{} \\ \hline
		\multirow{1}{*}{\begin{tabular}[c]{@{}c@{}}$\varpi^{e_{m,m'}}$ (Kbps)\end{tabular}} 
		& Between VMs & $10^4$ & $1.1\times10^5$ \\ \hline
		\multirow{1}{*}{\begin{tabular}[c]{@{}c@{}}$\tau^{e_{m,m'}}_{\max}$ (ms)\end{tabular}} & Between VMs & $5$ & $14$ \\ \hline
		\textbf{Other Parameters} & \multicolumn{3}{l|}{} \\ \hline
		\multirow{1}{*}{\begin{tabular}[c]{@{}c@{}}$|T|, |K|, |N|,$ and $|M|$\end{tabular}} 
		& \multicolumn{3}{l|}{1-16, 1, 4, and 3, respectively} \\ \hline
		\multirow{1}{*}{\begin{tabular}[c]{@{}c@{}}$\Upsilon$ and $\zeta$ \end{tabular}} 
		& \multicolumn{3}{l|}{1 and $9\times10^{-5}$} \\ \hline
	\end{tabular}
\end{table}
\setlength{\textfloatsep}{0pt}

\subsection{Simulation Results}
\begin{figure}[!h]
	\centering
	\includegraphics[width=\columnwidth]{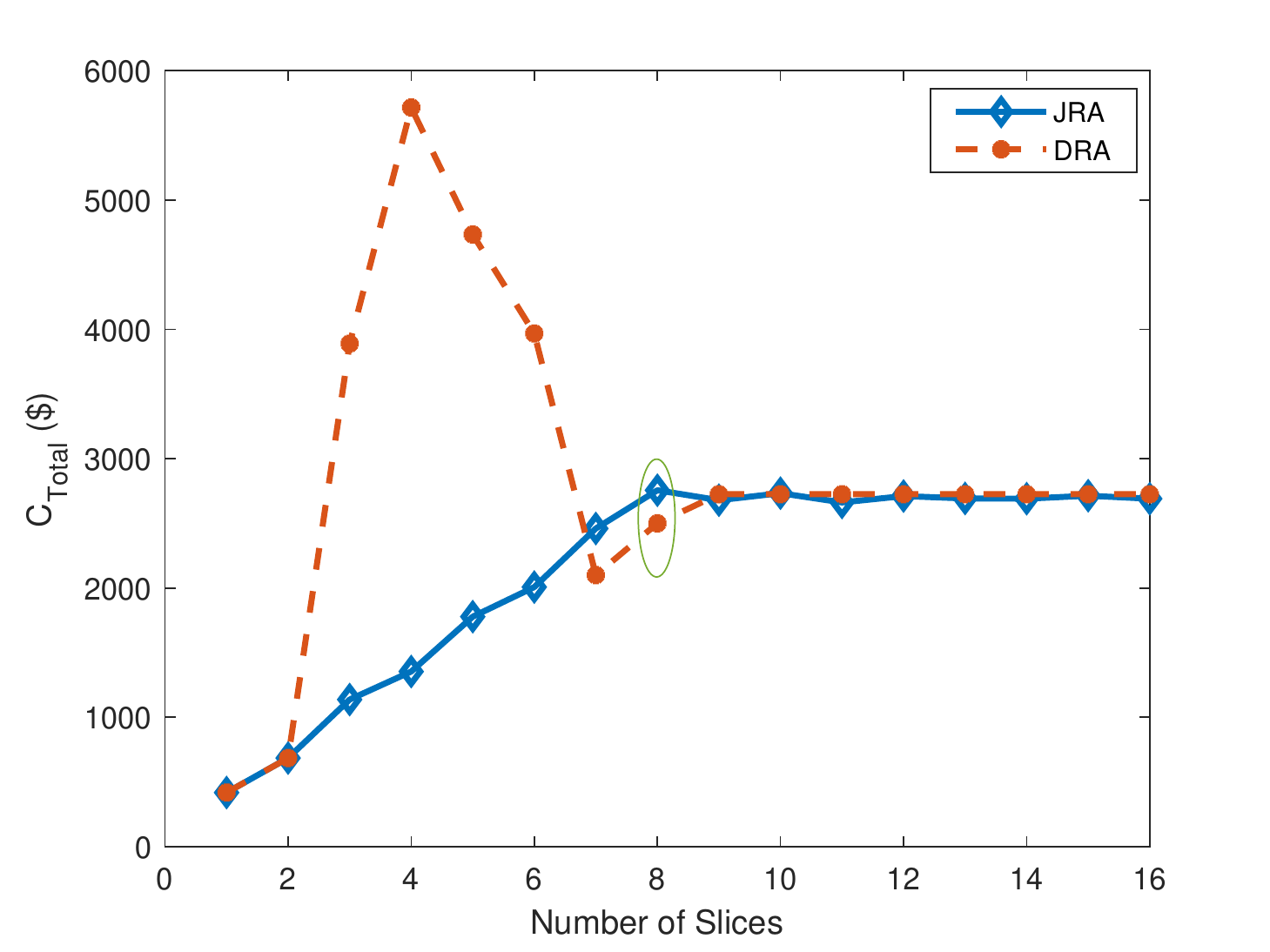}
	\caption{Comparison of the overall cost between the proposed methods.}\label{cost-comparison}
\end{figure}

\begin{figure}[!h]
	\centering
	\includegraphics[width=\columnwidth]{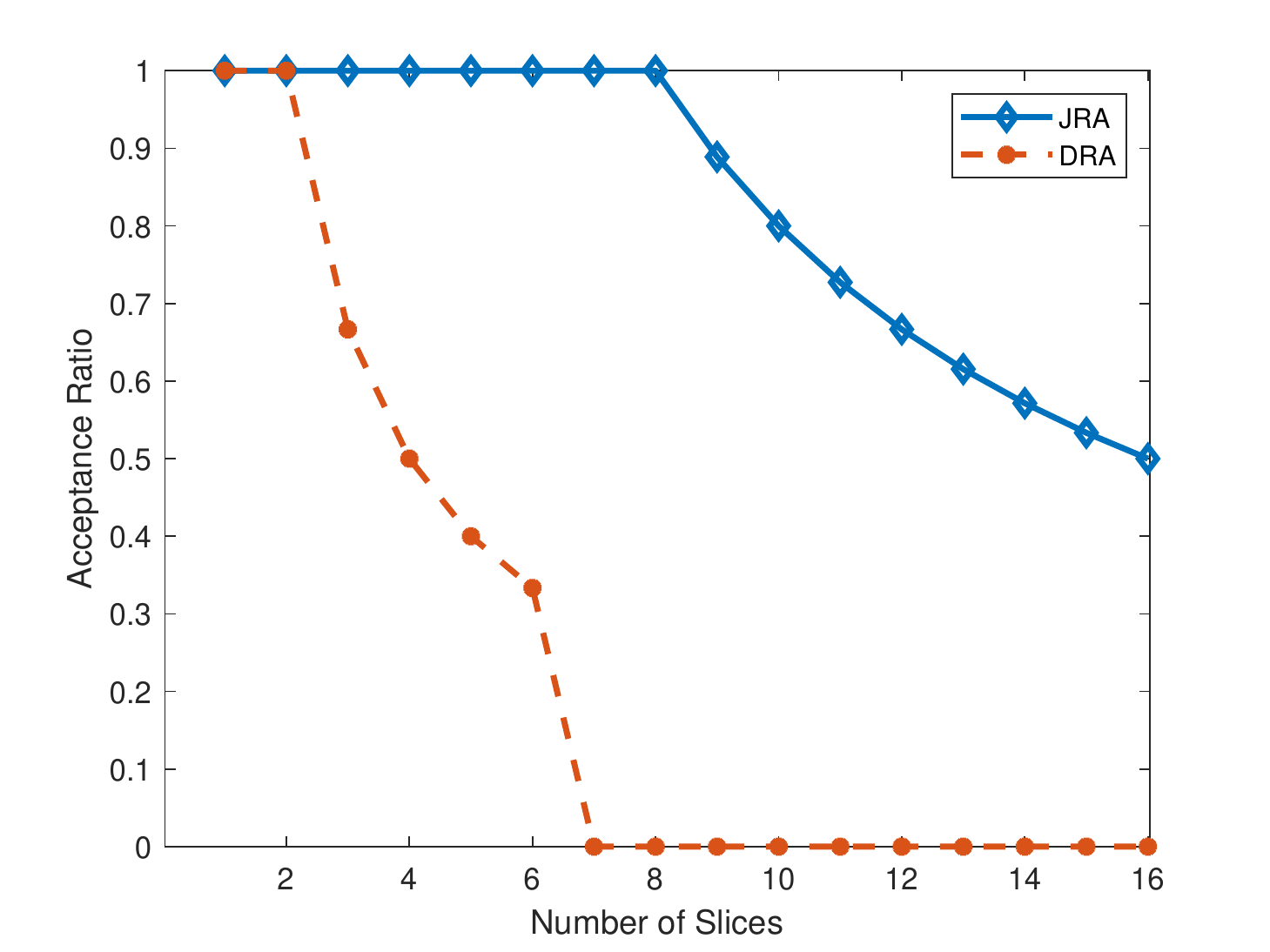}
	\caption{Comparison of the acceptance ratio between two proposed methods.}\label{acceptanceratio}
\end{figure}

\begin{figure}[!h]
	\centering
	\includegraphics[width=\columnwidth]{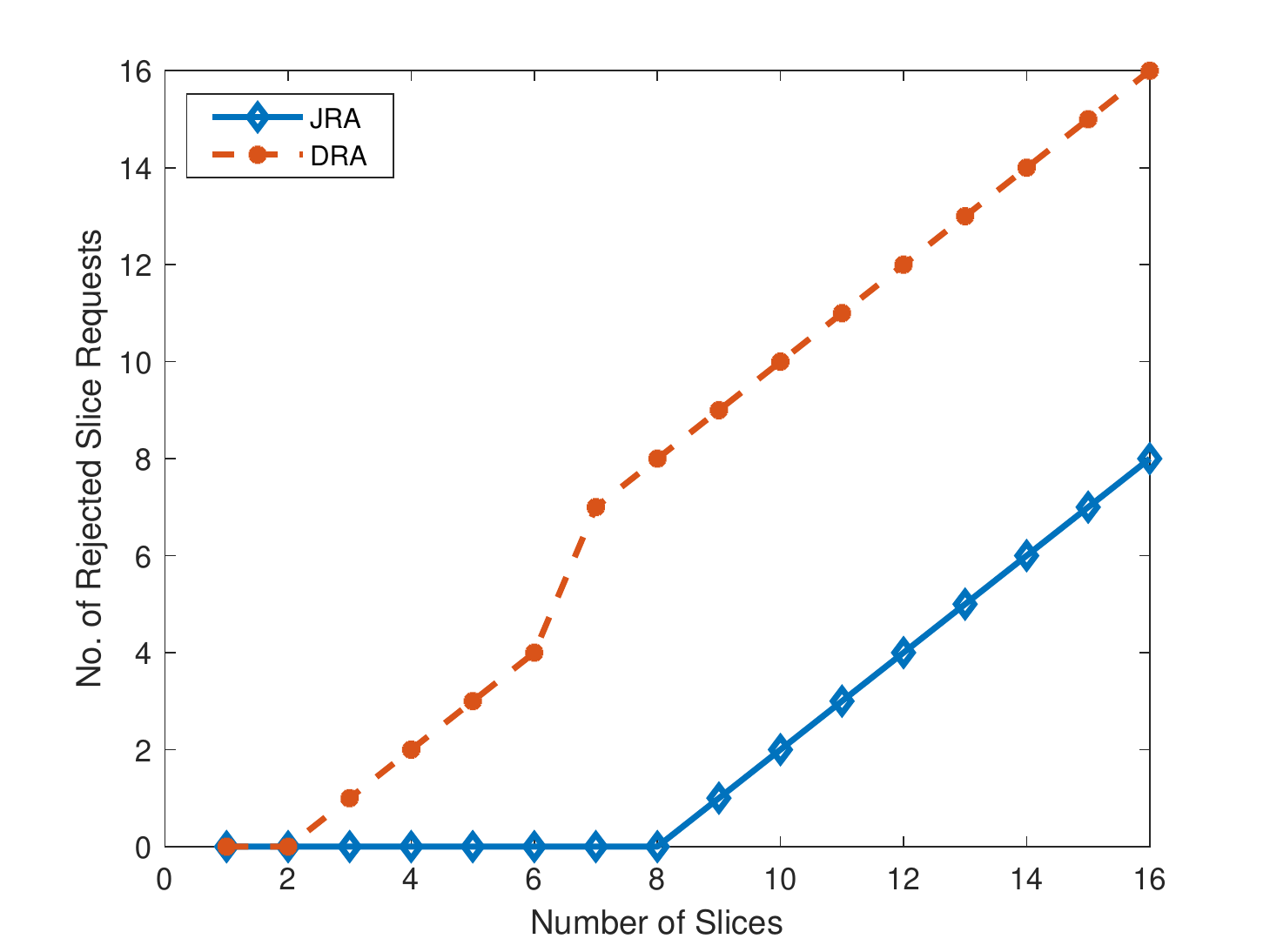}	
	\caption{Comparison of the number of {the} rejected slice requests between the proposed methods.}\label{rejectedslices}
\end{figure}

The overall cost (according to \eqref{objective_function}) of {the} CSP for serving the slice requests is shown in Fig. \ref{cost-comparison}. The {JRA method (i.e., Fig. \ref{flowchart})} is winning the race because the cost is growing smoothly using this method.\\
To analyze the performance of the admission control mechanisms (problem \eqref{admission_control} (i.e., AC-JRA) vs. \eqref{nodes_admission_control} (i.e., AC-DMA) and \eqref{links_admission_control} (i.e., AC-DLA)), Fig. \ref{acceptanceratio} displays the acceptance ratio of the two methods. {Hereby, the average difference between the acceptance ratio of JRA vs. DRA is 0.46, indicating the advantage of the JRA method. It is worth noting that} the DRA (i.e., Fig. \ref{disjoint-flowchart}) method, which is the intersection of AC-DMA and AC-DLA, cannot accept any slice request after the number of them reaches to $7$. This is the reason that the overall cost in Fig. \ref{cost-comparison} decreases dramatically, only showing the cost of subproblem \eqref{nodes_resource_allocation}. Moreover, Fig. \ref{rejectedslices} shows the number of the rejected slice requests in the two aforementioned methods.

\begin{figure}[!h]
	\centering
	\includegraphics[width=\columnwidth]{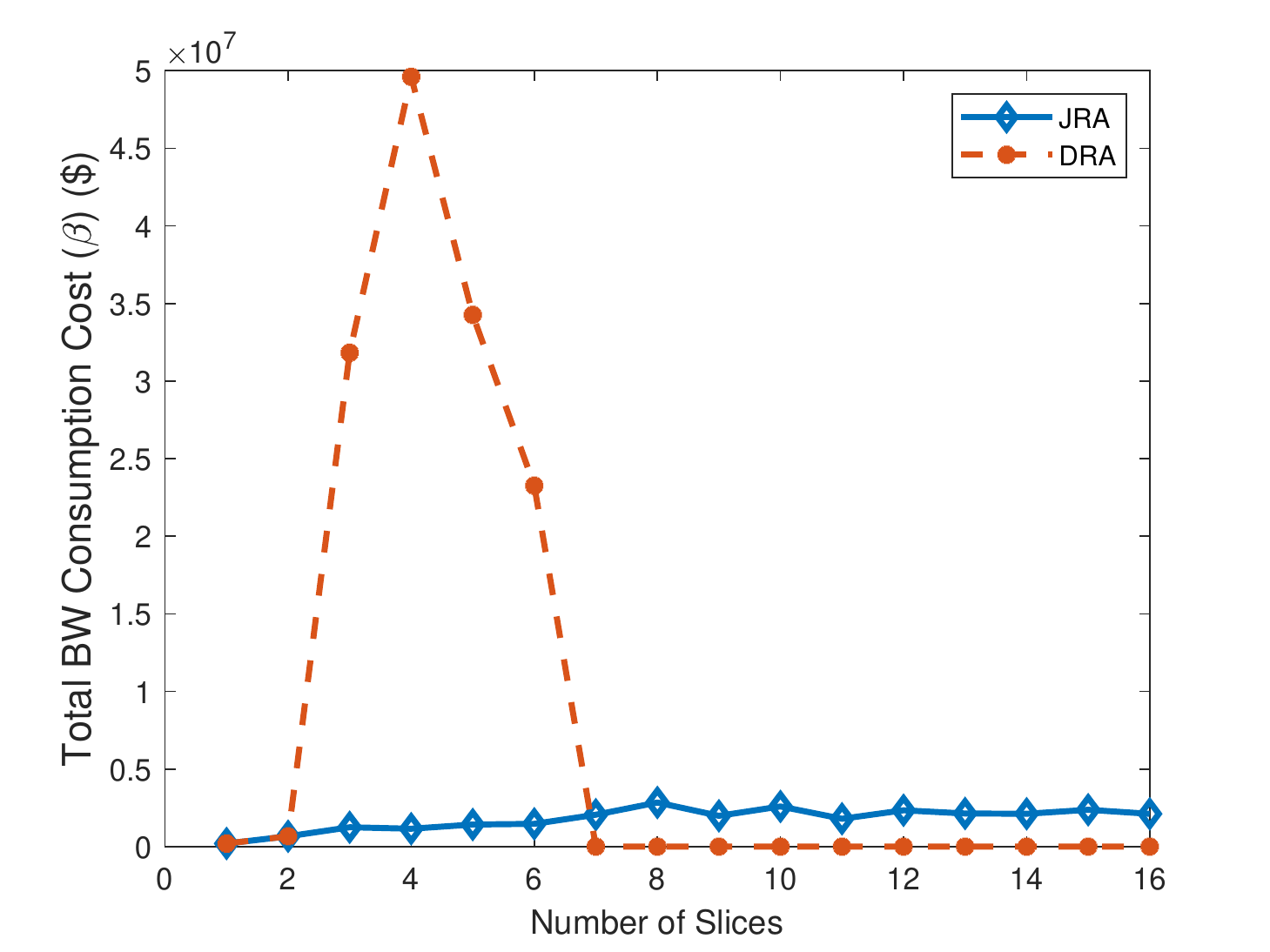}
	\caption{Comparison of the bandwidth consumption cost ($\beta$) between the proposed methods.}\label{bwconsumptionfig}
\end{figure}

Fig. \ref{bwconsumptionfig} shows the superiority of {the JRA method against the DRA method (i.e., Fig. \ref{disjoint-flowchart})} in terms of the bandwidth consumption cost. The $0$ value in {the DRA method} indicates that after the CSP gets $7$ slice requests, the AC-DLA subproblem (i.e., \eqref{links_admission_control}) is getting infeasible anyway and there will be no bandwidth consumption cost in \eqref{links_resource_allocation}. According to {Fig. \ref{acceptanceratio}}, after getting $7$ slice requests, computing resources are fulfilled and the CSP cannot accept any more slice requests, but in the DRA method, although after having $7$ slice requests, some requests will be accepted in AC-DMA subproblem (i.e., \eqref{nodes_admission_control}), the acceptance ratio of the AC-DLA subproblem (i.e., \eqref{links_admission_control}) suddenly drops to $0\%$. The main reason of this drop-off is that the links and nodes resources are related to each other and when we solve them in different phases, after getting some requests, subproblem \eqref{links_admission_control} will always be infeasible, thus leading to the cost of $0$ according to Fig. \ref{bwconsumptionfig}.

\begin{figure}[!h]
	\centering
	\includegraphics[width=\columnwidth]{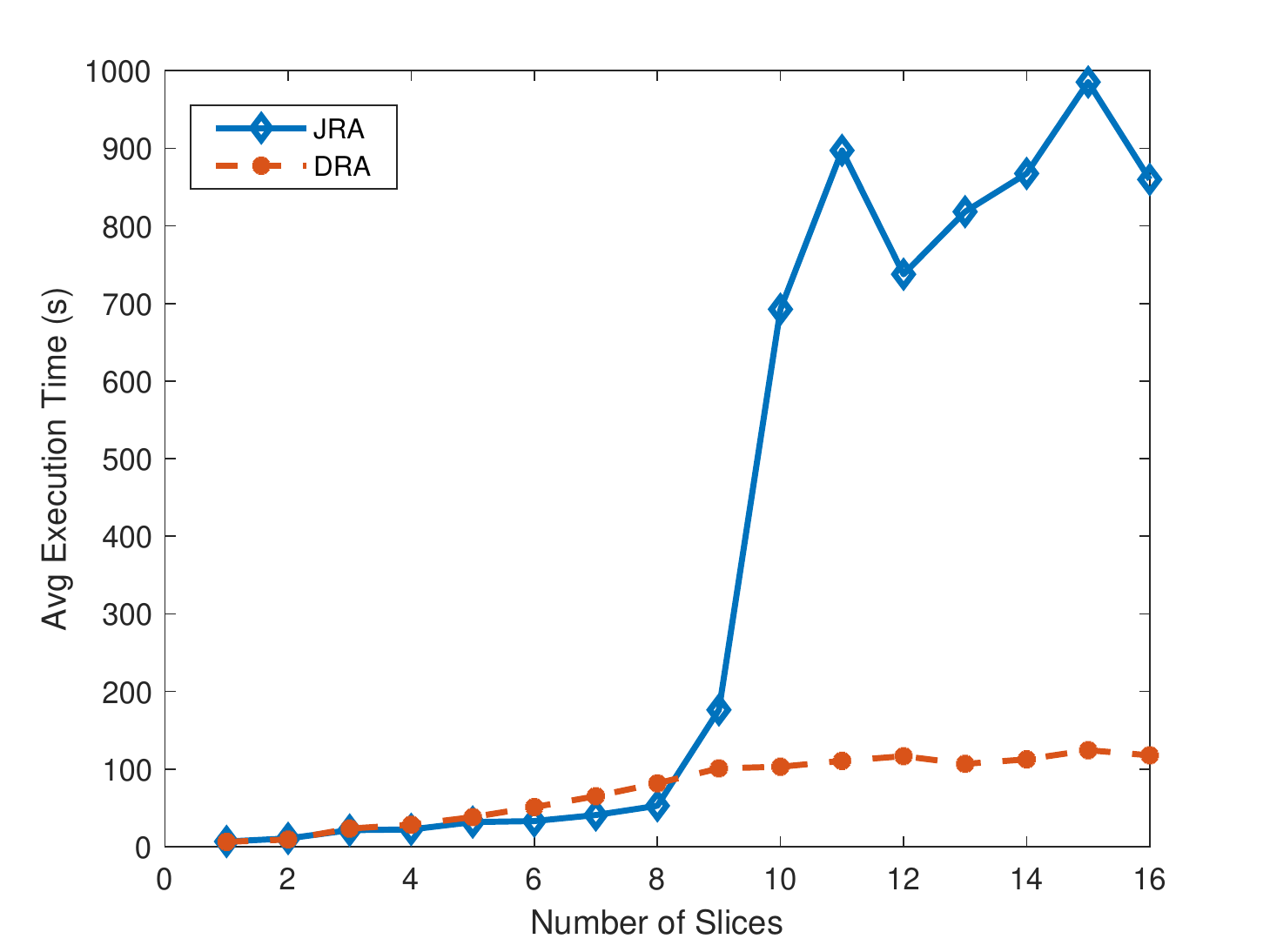}
	\caption{Comparison of average execution time between the proposed methods.}\label{executiontime}
\end{figure}

Fig. \ref{executiontime} presents the execution time of the JRA and DRA methods. As  can be seen, the JRA method has more execution time but according to Figs. \ref{cost-comparison}, \ref{acceptanceratio}, \ref{rejectedslices}, and \ref{bwconsumptionfig}, it outperforms the results of the DRA method significantly.

\section{Conclusion}\label{conclusion}
In this paper, we considered a network slicing problem where the CSP should allocate its resources according to the slice requests of its different tenants. An SDN/NFV-enabled network with some cloud nodes was considered and allocating the computing and bandwidth resources to different slices of CSP's tenants was the main problem. Because of the limited amount of resources, we performed an admission control mechanism before solving the joint problem to reject the slice requests which make the joint problem infeasible. To evaluate the joint method (i.e., Fig. \ref{flowchart}), we considered a new disjoint formulation (i.e., Fig. \ref{disjoint-flowchart}) of the original problem and the simulation results were all voted for our joint formulation.
{Incorporating the arrival time of each slice request into JRA and DRA can be regarded as a future line of research. Moreover, applying this system model in a discrete event simulator (e.g. NS3) will be done in our future work.}

%
\bibliographystyle{ieeetr}
\bibliography{citation_Slicing}{}

\begin{thebibliography}{10}

\bibitem{series2015imt}
M.~Series, ``{IMT} vision--framework and overall objectives of the future
  development of {IMT} for 2020 and beyond,'' {\em ITU Recommendation},
  pp.~2083--0, 2015.

\bibitem{alliance20155g}
{NGMN Alliance}, ``{5G} white paper,'' 2015.

\bibitem{alliance2016description}
{NGMN Alliance}, ``Description of network slicing concept,'' 2016.

\bibitem{rost2016mobile}
P.~Rost, A.~Banchs, I.~Berberana, M.~Breitbach, M.~Doll, H.~Droste,
  C.~Mannweiler, M.~A. Puente, K.~Samdanis, and B.~Sayadi, ``Mobile network
  architecture evolution toward {5G},'' {\em IEEE Communications Magazine},
  vol.~54, no.~5, pp.~84--91, 2016.

\bibitem{nikaein2015network}
N.~Nikaein, E.~Schiller, R.~Favraud, K.~Katsalis, D.~Stavropoulos, I.~Alyafawi,
  Z.~Zhao, T.~Braun, and T.~Korakis, ``Network store: Exploring slicing in
  future {5G} networks,'' in {\em Proc. of the 10th International Workshop on
  Mobility in the Evolving Internet Architecture (MobiArch 2015)}, pp.~8--13,
  ACM, 2015.

\bibitem{samdanis2016network}
K.~Samdanis, X.~Costa-Perez, and V.~Sciancalepore, ``From network sharing to
  multi-tenancy: The {5G} network slice broker,'' {\em IEEE Communications
  Magazine}, vol.~54, no.~7, pp.~32--39, 2016.

\bibitem{richart2016resource}
M.~Richart, J.~Baliosian, J.~Serrat, and J.-L. Gorricho, ``Resource slicing in
  virtual wireless networks: A survey,'' {\em IEEE Transactions on Network and
  Service Management}, vol.~13, no.~3, pp.~462--476, 2016.

\bibitem{zhang2017network}
H.~Zhang, N.~Liu, X.~Chu, K.~Long, A.-H. Aghvami, and V.~C. Leung, ``Network
  slicing based {5G} and future mobile networks: mobility, resource management,
  and challenges,'' {\em IEEE Communications Magazine}, vol.~55, no.~8,
  pp.~138--145, 2017.

\bibitem{afolabi2018network}
I.~Afolabi, T.~Taleb, K.~Samdanis, A.~Ksentini, and H.~Flinck, ``Network
  slicing and softwarization: A survey on principles, enabling technologies,
  and solutions,'' {\em IEEE Communications Surveys \& Tutorials}, vol.~20,
  no.~3, pp.~2429--2453, 2018.

\bibitem{ordonez2017network}
J.~Ordonez-Lucena, P.~Ameigeiras, D.~Lopez, J.~J. Ramos-Munoz, J.~Lorca, and
  J.~Folgueira, ``Network slicing for {5G} with {SDN}/{NFV}: Concepts,
  architectures, and challenges,'' {\em IEEE Communications Magazine}, vol.~55,
  no.~5, pp.~80--87, 2017.

\bibitem{su2019resource}
R.~Su, D.~Zhang, R.~Venkatesan, Z.~Gong, C.~Li, F.~Ding, F.~Jiang, and Z.~Zhu,
  ``Resource allocation for network slicing in {5G} telecommunication networks:
  A survey of principles and models,'' {\em IEEE Network (Early Access)},
  pp.~1--8, 2019.

\bibitem{sciancalepore2017slice}
V.~Sciancalepore, F.~Cirillo, and X.~Costa-Perez, ``Slice as a service
  ({SlaaS}) optimal {IoT} slice resources orchestration,'' in {\em Proc. of
  IEEE Global Communications Conference ({GLOBECOMM} 2017)}, pp.~1--7, IEEE,
  2017.

\bibitem{bega2017optimising}
D.~Bega, M.~Gramaglia, A.~Banchs, V.~Sciancalepore, K.~Samdanis, and
  X.~Costa-Perez, ``Optimising {5G} infrastructure markets: The business of
  network slicing,'' in {\em Proc. of IEEE International Conference on Computer
  Communications (INFOCOM 2017)}, pp.~1--9, IEEE, 2017.

\bibitem{han2018slice}
B.~Han, J.~Lianghai, and H.~D. Schotten, ``Slice as an evolutionary service:
  Genetic optimization for inter-slice resource management in {5G} networks,''
  {\em IEEE Access}, vol.~6, pp.~33137--33147, 2018.

\bibitem{leanh2016resource}
T.~LeAnh, N.~H. Tran, D.~T. Ngo, and C.~S. Hong, ``Resource allocation for
  virtualized wireless networks with backhaul constraints,'' {\em IEEE
  Communications Letters}, vol.~21, no.~1, pp.~148--151, 2016.

\bibitem{sattar2019optimal}
D.~Sattar and A.~Matrawy, ``Optimal slice allocation in {5G} core networks,''
  {\em IEEE Networking Letters}, vol.~1, no.~2, pp.~48--51, 2019.

\bibitem{fendt2018network}
A.~Fendt, S.~Lohmuller, L.~C. Schmelz, and B.~Bauer, ``A network slice resource
  allocation and optimization model for end-to-end mobile networks,'' in {\em
  Proc. of IEEE 5G World Forum (5GWF'18)}, pp.~262--267, IEEE, 2018.

\bibitem{wang2017resource}
G.~Wang, G.~Feng, W.~Tan, S.~Qin, R.~Wen, and S.~Sun, ``Resource allocation for
  network slices in {5G} with network resource pricing,'' in {\em Proc. of IEEE
  Global Communications Conference (GLOBECOMM 2017)}, pp.~1--6, IEEE, 2017.

\bibitem{kim2019reinforcement}
Y.~Kim, S.~Kim, and H.~Lim, ``Reinforcement learning based resource management
  for network slicing,'' {\em Applied Sciences}, vol.~9, no.~11, p.~2361, 2019.

\bibitem{caballero2018network}
P.~Caballero, A.~Banchs, G.~De~Veciana, X.~Costa-P{\'e}rez, and A.~Azcorra,
  ``Network slicing for guaranteed rate services: Admission control and
  resource allocation games,'' {\em IEEE Transactions on Wireless
  Communications}, vol.~17, no.~10, pp.~6419--6432, 2018.

\bibitem{5g2017view}
{5GPPP Architecture Working Group}, ``View on {5G} architecture, {Version
  2.0},'' {\em White Paper}, 2017.

\bibitem{zakeri2019energy}
A.~Zakeri, N.~Gholipoor, M.~R. Javan, and N.~Mokari, ``Energy cost minimization
  by joint radio and {NFV} resource allocation: {E}2{E} {Q}o{S} framework,''
  {\em arXiv preprint arXiv:1907.06212}, 2019.

\bibitem{gao2013quality}
Y.~Gao, H.~Guan, Z.~Qi, B.~Wang, and L.~Liu, ``Quality of service aware power
  management for virtualized data centers,'' {\em Journal of Systems
  Architecture}, vol.~59, no.~4-5, pp.~245--259, 2013.

\bibitem{beloglazov2012energy}
A.~Beloglazov, J.~Abawajy, and R.~Buyya, ``Energy-aware resource allocation
  heuristics for efficient management of data centers for cloud computing,''
  {\em Future generation computer systems}, vol.~28, no.~5, pp.~755--768, 2012.

\bibitem{dayarathna2016data}
M.~Dayarathna, Y.~Wen, and R.~Fan, ``Data center energy consumption modeling: A
  survey,'' {\em IEEE Communications Surveys \& Tutorials}, vol.~18, no.~1,
  pp.~732--794, 2016.

\bibitem{addad2018towards}
R.~A. Addad, T.~Taleb, M.~Bagaa, D.~L.~C. Dutra, and H.~Flinck, ``Towards
  modeling cross-domain network slices for {5G},'' in {\em Proc. of IEEE Global
  Communications Conference (GLOBECOM 2018)}, pp.~1--7, IEEE, 2018.

\bibitem{Chinneck2008feasibility}
J.~W. Chinneck, {\em Feasibility and Infeasibility in Optimization}.
\newblock Algorithms and Computational Methods, Springer, 2008.

\bibitem{tajallifar2019qosaware}
M.~Tajallifar, S.~Ebrahimi, M.~R. Javan, and N.~Mokari, ``{QoS}-aware joint
  power allocation and task offloading in an {MEC}/{NFV}-enabled {C-RAN}
  network,'' {\em arXiv preprint arXiv:1912.00187}, 2019.

\bibitem{mosek}
{MOSEK ApS}, {\em The MOSEK optimization toolbox for MATLAB manual. Version
  9.0.}, 2019.

\bibitem{cvx}
{CVX Research, Inc.}, ``{CVX}: Matlab software for disciplined convex
  programming, {Version} 2.0,'' 2012.

\bibitem{matlab}
{MathWorks}, {\em {MATLAB}}, 2019.

\bibitem{tajiki2019joint}
M.~M. Tajiki, S.~Salsano, L.~Chiaraviglio, M.~Shojafar, and B.~Akbari, ``Joint
  energy efficient and {QoS}-aware path allocation and {VNF} placement for
  service function chaining,'' {\em IEEE Transactions on Network and Service
  Management}, vol.~16, no.~1, pp.~374--388, 2019.

\end{thebibliography}
\end{document}